\documentclass{article}
\usepackage{spconf,amsmath,amssymb,graphicx,booktabs,multirow,url}
\usepackage{tikz,pgfplots}
\pgfplotsset{compat=1.16}
\usetikzlibrary{positioning,arrows.meta,shapes.geometric,decorations.pathmorphing,calc,fit,backgrounds}
\definecolor{ccqfm}{HTML}{0072B2}
\definecolor{cprior}{HTML}{009E73}
\definecolor{chaar}{HTML}{E69F00}
\definecolor{cquddpm}{HTML}{D55E00}

\def\x{{\mathbf x}}

\title{Conditional Quantum Flow Matching for Data-Scarce Physiological Signal Augmentation}

\name{Chi-Sheng Chen$^{\dagger}$ \qquad Samuel Yen-Chi Chen$^{\star}$}
\address{$^{\dagger}$Harvard Medical School \& Beth Israel Deaconess Medical Center, Boston, MA, USA\\
         $^{\star}$Brookhaven National Laboratory, Upton, NY, USA\\
         \texttt{m50816m50816@gmail.com}, \texttt{ycchen1989@ieee.org}\\
         {\footnotesize \textsc{orcid}\, 0000-0003-0807-0217 \quad
          0000-0003-0114-4826}}

\begin{document}
\maketitle

\begin{abstract}
Generative augmentation is a standard remedy for label scarcity in
physiological signal classification, but existing quantum generative
models start from uninformative noise, ignoring class structure that
is already available. We propose Conditional Quantum
Flow Matching (CQFM): a single 306-parameter circuit, conditioned on
both flow time and class label, transports a compact class-conditional
prior toward the target distribution.
Quantum flow matching as published is unconditional, so this is to
our knowledge the first conditional one, and the first EEG
augmentation on a parameterized quantum circuit. A nonnegative
spectral embedding removes the need for tomography at readout. On BCI Competition IV-2a, starting from a prior rather
than noise is worth $+5.1$ accuracy points over QuDDPM (9/9 subjects),
though at that operating point a class-conditional Gaussian matches
CQFM. Where the prior fails the transport earns its keep: given one
transferred from other subjects it regains $+7.2$ TSTR points (9/9).
\end{abstract}

\begin{keywords}
quantum machine learning, flow matching, generative models, EEG, data augmentation
\end{keywords}

\section{Introduction}
\label{sec:intro}

Deep classifiers for physiological signals such as EEG rarely have
enough labels: trials are costly to record, subjects fatigue, and
distributions shift across sessions. Generative augmentation is a
common remedy, and quantum generative models have recently been
applied to it. Quantum GANs synthesize ECG beats
\cite{qcganecg2023,hqdcgan2024}, parameterized quantum circuits
refine latent diffusion for arrhythmia augmentation
\cite{kritopoulos2026quantum}, and quantum diffusion models generate
financial time series \cite{qdiffusionts2026} and images
\cite{chen2026qrldiff}. All of these start
their transport from uninformative noise, the quantum analogue of
starting a flow at a standard Gaussian. Quantum machine learning has
reached EEG for encoding and representation learning
\cite{chen2024qeegnet,chen2025qcl}, and quantum-\emph{inspired} GANs
have synthesized EEG for privacy \cite{qdpgan2026}, but not, to our
knowledge, an EEG generator on a parameterized quantum circuit.

Quantum flow matching (QFM) \cite{cui2025qfm} extends classical flow
matching \cite{lipman2023flow} to quantum ensembles: a circuit of
alternating unitary and partly-measured layers interpolates between
two \emph{arbitrary} density matrices. So far QFM has only been
applied to quantum systems themselves, for example Ising thermal
states and Jarzynski estimates, and it is \emph{unconditional}: one
trained circuit carries one source ensemble to one target, so
generating labelled data would need a separate model per class.

Two things are therefore missing for our purpose. The first is a
conditioning mechanism. The second is that QFM's source ensemble does
not have to be Haar-random---the freedom that noise-initialized
generative models leave unused. In the low-label setting we can build
a compact class-conditional prior from the few labeled trials and ask
a single conditioned circuit to learn only the remaining transport.

\textbf{Contributions.} (i)~\emph{A conditional QFM model}
(Sec.~\ref{sec:method}). QFM as published has no conditioning
mechanism: one trained circuit carries one ensemble to one target.
CQFM folds the flow time and the class label into the same rotation
angles, so a single 306-parameter circuit realizes every transport
step of every class, trained end-to-end with a differentiable
branch-averaged fidelity loss on minibatch-OT-paired interpolants. To
our knowledge this is the first \emph{conditional} quantum
flow-matching model, the first applied to classical signals, and the
first EEG augmentation method built on a parameterized quantum
circuit. An ablation locates the gain in the conditional orthogonal
parameterization rather than in the quantum channel. (ii)~\emph{Encoding
design} (Sec.~\ref{ssec:encoding}). A real-orthogonal ansatz is
paired with a nonnegative spectral-magnitude amplitude embedding and
class-discriminative bin selection, so generated states can be read
out without tomography. (iii)~\emph{Experiments}
(Sec.~\ref{sec:exp}). Comparisons against the official QuDDPM
\cite{zhang2024quddpm}, a prior-swap ablation, and tuned classical
generators show that the informative prior, not circuit depth,
determines generation quality, and that the results hold down to 1024
measurement shots.

\section{Method}
\label{sec:method}

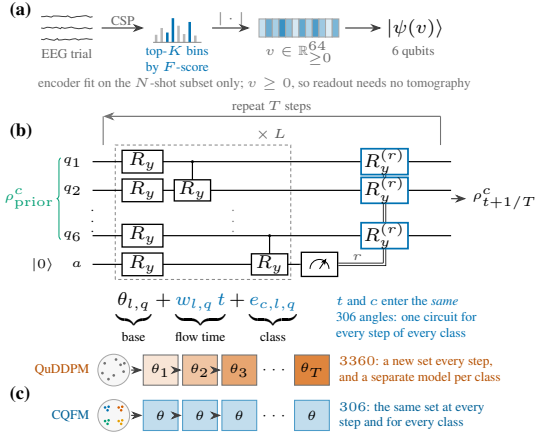
\begin{figure}[t]
\centering
\begin{tikzpicture}[
  font=\scriptsize,
  wire/.style={line width=0.45pt, black},
  gate/.style={draw, line width=0.5pt, fill=white, inner sep=0pt,
               minimum height=4.4mm, minimum width=6.8mm, font=\scriptsize},
  cond/.style={gate, draw=ccqfm, line width=0.65pt},
  cwire/.style={line width=0.3pt, black, double, double distance=0.65pt},
  arr/.style={-{Stealth[length=1.5mm,width=1.1mm]}, line width=0.45pt, black!75},
  tag/.style={font=\tiny, text=black!60, inner sep=1pt},
  panel/.style={font=\scriptsize\bfseries},
  stp/.style={draw=black!55, line width=0.4pt, minimum width=4.6mm,
              minimum height=4.0mm, inner sep=0pt, font=\tiny},
]

\def\ay{0}
\node[panel, anchor=west] at (-0.10,\ay+0.54) {(a)};

\begin{scope}[shift={(0.44,\ay)}]
  \foreach \i in {0,1,2}
    \draw[line width=0.3pt, black!75, decorate,
          decoration={random steps, segment length=1.15pt, amplitude=0.42pt}]
      (0,{0.46-0.15*\i}) -- ++(0.68,0);
  \node[tag, anchor=north] at (0.34,0.07) {EEG trial};
\end{scope}

\draw[arr] (1.24,\ay+0.31) -- node[tag, above, inner sep=0.8pt] {CSP} ++(0.56,0);

\begin{scope}[shift={(1.96,\ay+0.12)}]
  \foreach \i/\h/\s in {0/.10/0, 1/.19/0, 2/.13/1, 3/.32/1, 4/.22/0,
                        5/.11/0, 6/.27/1, 7/.09/0}
    \draw[line width=0.95pt, color={\ifnum\s=1 ccqfm\else black!28\fi}]
      ({0.074*\i},0) -- ++(0,\h);
  \draw[line width=0.3pt, black!45] (-0.04,0) -- (0.60,0);
  \node[tag, anchor=north, text=ccqfm, align=center] at (0.28,-0.01)
    {top-$K$ bins\\by $F$-score};
\end{scope}

\draw[arr] (2.70,\ay+0.31) -- node[tag, above, inner sep=0.8pt] {$|\cdot|$} ++(0.50,0);

\begin{scope}[shift={(3.32,\ay+0.19)}]
  \foreach \i/\v in {0/22, 1/58, 2/34, 3/78, 4/45, 5/16, 6/62, 7/29,
                     8/70, 9/38, 10/13, 11/52}
    \fill[ccqfm!\v!white, draw=black!35, line width=0.2pt]
      ({0.088*\i},0) rectangle ++(0.088,0.24);
  \node[tag, anchor=north] at (0.53,-0.01) {$v\in\mathbb{R}^{64}_{\ge 0}$};
\end{scope}

\draw[arr] (4.46,\ay+0.31) -- ++(0.48,0);
\node[anchor=west, inner sep=0pt] at (5.00,\ay+0.31) {$|\psi(v)\rangle$};
\node[tag, anchor=north] at (5.36,\ay+0.11) {6 qubits};

\node[tag, anchor=north west, text=black!55] at (0.36,\ay-0.30)
  {encoder fit on the $N$-shot subset only; $v\ge 0$, so readout needs no
   tomography};

\def\by{-2.44}
\node[panel, anchor=west] at (-0.10,\by+1.38) {(b)};

\def\ra{\by+0.98}
\def\rb{\by+0.60}
\def\rdots{\by+0.30}
\def\rd{\by+0.00}
\def\re{\by-0.38}

\foreach \y/\lb in {\ra/{$q_1$}, \rb/{$q_2$}, \rd/{$q_6$}, \re/{$a$}}
  \node[anchor=east, inner sep=1.4pt, font=\tiny] at (1.04,\y) {\lb};
\node[anchor=east, inner sep=0pt, text=cprior, font=\tiny] at (0.62,\by+0.49)
  {$\rho^{c}_{\mathrm{prior}}$};
\node[anchor=east, inner sep=0pt, font=\tiny] at (0.62,\re) {$|0\rangle$};
\draw[decorate, decoration={brace, amplitude=1.4pt}, line width=0.3pt, cprior]
  (0.70,\rd-0.05) -- (0.70,\ra+0.05);

\foreach \y in {\ra,\rb,\rd} \draw[wire] (1.12,\y) -- (5.86,\y);
\draw[wire] (1.12,\re) -- (3.86,\re);
\node[font=\tiny, inner sep=0pt] at (1.12,\rdots) {$\vdots$};

\foreach \y in {\ra,\rb,\rd,\re}
  \node[gate, minimum width=5.4mm, minimum height=3.0mm] at (1.78,\y) {$R_y$};

\draw[wire] (2.44,\ra) -- (2.44,\rb);
\fill (2.44,\ra) circle (0.55pt);
\node[gate, minimum width=4.8mm, minimum height=2.8mm] at (2.44,\rb) {$R_y$};
\node[font=\tiny, inner sep=0pt, text=black!55] at (2.98,\rdots) {$\vdots$};
\draw[wire] (3.46,\rd) -- (3.46,\re);
\fill (3.46,\rd) circle (0.55pt);
\node[gate, minimum width=4.8mm, minimum height=2.8mm] at (3.46,\re) {$R_y$};

\begin{scope}[on background layer]
  \draw[draw=black!45, line width=0.4pt, dash pattern=on 1.2pt off 1.0pt]
    (1.42,\by-0.58) rectangle (3.74,\by+1.20);
\end{scope}
\node[tag, anchor=south east, text=black!65] at (3.72,\by+1.22) {$\times\,L$};

\node[gate, minimum width=4.8mm, minimum height=3.2mm] (mz) at (4.12,\re) {};
\draw[line width=0.3pt] ([xshift=-1.25mm,yshift=-0.6mm]mz.center)
  arc[start angle=180, end angle=0, radius=1.25mm];
\draw[line width=0.3pt, -{Stealth[length=0.8mm,width=0.6mm]}]
  ([yshift=-0.6mm]mz.center) -- ++(0.9mm,1.35mm);
\draw[wire] (3.86,\re) -- (mz.west);
\node[tag, anchor=south, text=black!65, inner sep=1.3pt] at (4.62,\re+0.01) {$r$};

\draw[cwire] (mz.east) -- (4.98,\re) -- (4.98,\rd-0.16);
\draw[cwire] (4.98,\rd+0.15) -- (4.98,\rb-0.15);
\draw[cwire] (4.98,\rb+0.15) -- (4.98,\ra-0.15);
\foreach \y in {\ra,\rb,\rd}
  \node[cond, minimum width=6.2mm, minimum height=3.0mm] at (4.98,\y)
    {$R_y^{(r)}$};

\draw[arr] (5.86,\by+0.49) -- ++(0.20,0);
\node[anchor=west, inner sep=1pt, font=\tiny] at (6.10,\by+0.49)
  {$\rho^{c}_{t+1/T}$};

\draw[line width=0.4pt, black!55] (5.72,\ra+0.22) -- (5.72,\by+1.58);
\draw[line width=0.4pt, black!55] (1.26,\by+1.58) -- (1.26,\ra+0.22);
\draw[arr, black!55] (5.72,\by+1.58)
  -- node[tag, above, inner sep=1.2pt, text=black!60] {repeat $T$ steps}
  (1.26,\by+1.58);

\node[anchor=north west, inner sep=0pt] at (1.42,\by-0.74)
  {$\underbrace{\theta_{l,q}}_{\text{\tiny base}}
    +\underbrace{\textcolor{ccqfm}{w_{l,q}\,t}}_{\text{\tiny flow time}}
    +\underbrace{\textcolor{ccqfm}{e_{c,l,q}}}_{\text{\tiny class}}$};
\node[tag, anchor=north west, text=ccqfm, align=left] at (4.30,\by-0.76)
  {$t$ and $c$ enter the \emph{same}\\306 angles: one circuit for\\
   every step of every class};
\def\cy{-4.46}
\node[panel, anchor=west] at (-0.10,\cy-0.08) {(c)};

\def\qa{\cy+0.24}
\def\qb{\cy-0.36}

\node[tag, anchor=east, text=cquddpm!85!black] at (1.12,\qa) {QuDDPM};
\node[tag, anchor=east, text=ccqfm!85!black]   at (1.12,\qb) {CQFM};

\begin{scope}[shift={(1.40,\qa)}]
  \draw[draw=black!35, line width=0.35pt, fill=black!3] (0,0) circle (0.21);
  \foreach \x/\y in {-.12/.09, .03/.14, .13/-.02, -.07/-.12, .09/.07,
                     -.15/-.04, .00/-.07, .14/.09}
    \fill[black!55] (\x,\y) circle (0.018);
\end{scope}
\begin{scope}[shift={(1.40,\qb)}]
  \draw[draw=black!35, line width=0.35pt, fill=black!3] (0,0) circle (0.21);
  \foreach \cx/\vy/\c in {-.09/.09/ccqfm, .09/.09/cquddpm,
                          -.09/-.09/cprior, .09/-.09/chaar}
    \foreach \dx/\dy in {-.025/.015, .022/.025, .008/-.022}
      \fill[\c] ({\cx+\dx},{\vy+\dy}) circle (0.017);
\end{scope}

\foreach \lb/\x/\sh in {{$\theta_1$}/2.02/16, {$\theta_2$}/2.54/36,
                        {$\theta_3$}/3.06/56, {$\theta_T$}/4.02/76}
  \node[stp, fill=cquddpm!\sh!white] at (\x,\qa) {\lb};
\node[font=\tiny] at (3.54,\qa) {$\cdots$};
\foreach \x in {2.02, 2.54, 3.06, 4.02}
  \node[stp, fill=ccqfm!24!white, draw=ccqfm!70] at (\x,\qb) {$\theta$};
\node[font=\tiny] at (3.54,\qb) {$\cdots$};

\foreach \y in {\qa,\qb} {
  \draw[arr] (1.63,\y) -- (1.78,\y);
  \foreach \a/\b in {2.02/2.54, 2.54/3.06} \draw[arr] ({\a+0.24},\y) -- ({\b-0.24},\y);
}

\node[tag, anchor=west, align=left, text=cquddpm!85!black] at (4.34,\qa)
  {$3360$: a new set every step,\\and a separate model per class};
\node[tag, anchor=west, align=left, text=ccqfm!85!black] at (4.34,\qb)
  {$306$: the same set at every\\step and for every class};
\end{tikzpicture}
\caption{CQFM. (a)~Encoding: CSP components, the $K$ most
class-discriminative rFFT bins each, magnitudes forming a nonnegative
64-dim vector. (b)~One flow step on six system qubits and an ancilla:
$L$ layers of $R_y$ and ring-$CR_y$ (ring closed by $a\!\to\!q_1$;
$\vdots$ = elided couplings), then the ancilla is measured and $r$
selects an $R_y$ correction, making the step a channel. (c)~QuDDPM
starts at a Haar state with a fresh angle set per step and class
($3360$).}
\label{fig:method}
\end{figure}

\subsection{Spectral encoding}
\label{ssec:encoding}

Let a labeled trial be $X \in \mathbb{R}^{C\times T_s}$ with label
$y$. We fit, \emph{on the $N$-shot labeled subset only}: (i) CSP
spatial filters \cite{ramoser2000csp} $W \in \mathbb{R}^{4\times C}$
(csp-space projection); (ii) per-component masks selecting the $K{=}16$
rFFT bins with the highest ANOVA $F$-score of spectral magnitude
against the class label. The encoding is the concatenated magnitude
spectrum $v = [\,|F_1|_{m_1},\dots,|F_4|_{m_4}\,] \in
\mathbb{R}^{64}_{\ge 0}$, $\ell_2$-normalized and prepared as an
amplitude-encoded state $|\psi(v)\rangle = \sum_j v_j |j\rangle$ on
$n{=}6$ qubits. This is \emph{feature-space} augmentation: the
magnitude encoding cannot be inverted back to raw EEG, and all
downstream evaluation takes place in the same space. Two design choices matter.
Complex spectra did not learn at all (0.258 across nine
subjects, against 0.25 chance); discarding phase reaches 0.389, so
single-trial spectral phase is mostly noise here. And keeping amplitudes near-nonnegative removes the
need for tomography at readout. The real-orthogonal transport does not
preserve nonnegativity exactly, but because it starts near identity
from nonnegative inputs, the negative-amplitude probability mass
stays below $0.2\%$ in our measurements, and the $v_j = \sqrt{p_j}$
computational-basis readout removes the remainder.

\subsection{Class-conditional prior}

For each class $c$ we fit a diagonal Gaussian
$\mathcal{N}(\mu_c,\sigma_c^2)$ to the encoded subset vectors and take
the normalized samples as the prior ensemble
$\rho^c_{\mathrm{prior}}$. This is the informative source that QFM
allows and that noise-initialized models do not use.

\subsection{The CQFM model}

QFM \cite{cui2025qfm} provides the transport primitive: circuits with
measured ancillas can interpolate between arbitrary ensembles. The
conditioning mechanism, training objective, prior construction, and
encoding design described here are new. Each flow step is a channel
rather than a unitary. The system state and one ancilla pass through
$L$ layers of $R_y$ rotations and ring-$CR_y$ entanglers (a
real-orthogonal family, so states remain real, and the map is close
to the identity at small angles); the ancilla is then measured, and
an $R_y$ correction conditioned on the outcome is applied to the
system. The rotation angles are
$\theta_{l,q} + w_{l,q}\,t + e_{c,l,q}$, so the flow time $t$ and
class label $c$ enter a \emph{single} shared circuit: 306 parameters
in total, namely 42 base angles $\theta$, 42 time weights $w$, 168
class embeddings $e$ (4 classes), 42 entangler angles, and 12
outcome-conditioned corrections. A fixed CNOT ring was tried and
discarded: it is a basis permutation and destroys the near-identity
initialization small-step flows need. The measured ancilla, by
contrast, turns out not to be load-bearing (Sec.~\ref{sec:exp}).

Training samples a class $c$, a prior point $x_0$, a data point
$x_1$, and a step $k$; the circuit at $t = k/T$ is trained to map the
normalized interpolant $x_t \propto (1{-}t)x_0 + t x_1$ to
$x_{t+1/T}$. This is a discrete-time analogue of conditional flow
matching \cite{lipman2023flow}: a per-step map regression along
OT-paired interpolant paths, without the marginal guarantees of a
continuous vector field, so distributional quality is checked
empirically in Sec.~\ref{sec:exp}. The loss is the branch-averaged
infidelity
$\sum_r p_r\,(1 - \langle \psi_r | \psi_{\mathrm{tgt}}\rangle^2)$,
which can be differentiated without sampling; at generation time the
ancilla outcomes are sampled. Pairs $(x_0, x_1)$ are matched by
minibatch optimal transport \cite{tong2024improving}, which reduced
the loss floor from 0.025 to 0.0175. We use $T{=}10$ steps, $L{=}6$
layers, and 1500 Adam steps.

\section{Experiments}
\label{sec:exp}

\textbf{Setup.} BCI Competition IV-2a \cite{tangermann2012review}
(9 subjects, 4-class motor imagery, 22 channels): session T for
training, session E held out. Low-label protocol: $N \in
\{10,25,50,72\}$ labeled trials per class, 5 subsampling seeds; the
encoder, priors, generative models, and classifiers see only the
$N$-shot subset. Baselines: no augmentation; the informative prior
alone (no transport); CQFM with a Haar prior (prior-swap ablation,
identical circuit and budget); the \emph{official} QuDDPM
implementation \cite{zhang2024quddpm} (per-class models, $T{=}10$,
$L{=}6$, 3360 parameters total vs.~CQFM's 306;
Fig.~\ref{fig:method}c); and \emph{classical
conditional generators in the identical feature space}: a
class-conditional Ledoit-Wolf Gaussian, a conditional VAE, and
classical conditional flow matching \cite{lipman2023flow,
tong2024improving} with an MLP velocity field at two budgets
(332 and 33k parameters), trained with the same prior, OT pairing,
steps, and 10-step Euler sampling. Integer hidden widths cannot hit 306
exactly (width 1 gives 198, width 2 gives 332), so the $\alpha$ study
below brackets CQFM from both sides. Metrics: augmented
downstream accuracy of an MLP (real $N$ + 800 synthetic), C2ST
\cite{lopezpaz2017c2st} real-vs-generated discriminability
(logistic-regression 5-fold CV; 0.5 = indistinguishable), and
class-mean log-spectral distance (LSD, dB). Statistics: paired Wilcoxon
with \emph{subjects} as the unit ($n{=}9$, seed- and budget-averaged;
per-configuration analyses agree in direction). Circuits are simulated
exactly (PennyLane 0.38 statevector, PyTorch autograd, CPU); seeds
$0$--$4$ are shared across methods. Code, configurations, and all summary results
are at \url{https://github.com/ChiShengChen/cqfm-icassp2027}.

\begin{table}[t]
\centering
\caption{Augmented downstream MLP accuracy (mean over 9 subjects
$\times$ 5 seeds) and distributional fidelity at $N{=}25$ (C2ST and
LSD: lower is better). cFM rows use each budget's best of four
hyperparameter settings, selected in the classical model's favor.
Bold: best per column.}
\label{tab:main}
\footnotesize
\setlength{\tabcolsep}{3.6pt}
\begin{tabular}{lcccccc}
\toprule
& \multicolumn{4}{c}{Accuracy ($N$ per class)}
& \multicolumn{2}{c}{Fidelity ($N{=}25$)} \\
\cmidrule(lr){2-5} \cmidrule(lr){6-7}
Method & 10 & 25 & 50 & 72 & C2ST & LSD \\
\midrule
No augmentation      & .353 & .385 & .392 & .399 & -- & -- \\
cVAE                 & .358 & .395 & .390 & .384 & .915 & 1.50 \\
cGauss (Ledoit-Wolf) & .359 & .390 & .405 & .408 & .832 & 2.40 \\
cFM (332 params)     & .370 & .405 & .420 & .428 & .813 & 1.67 \\
cFM (33k params)     & \textbf{.372} & .408 & \textbf{.426} & .426
                     & \textbf{.799} & \textbf{1.42} \\
Prior only           & .371 & .404 & .421 & .429 & .800 & 1.43 \\
\midrule
CQFM (Haar prior)    & .310 & .355 & .373 & .376 & .982 & 5.20 \\
QuDDPM (official)    & .313 & .362 & .368 & .381 & .982 & 5.16 \\
\textbf{CQFM (informative)} & .368 & \textbf{.411} & .418
                     & \textbf{.432} & .800 & 1.45 \\
\bottomrule
\end{tabular}
\end{table}

\begin{figure}[t]
\centering
\includegraphics[width=\columnwidth]{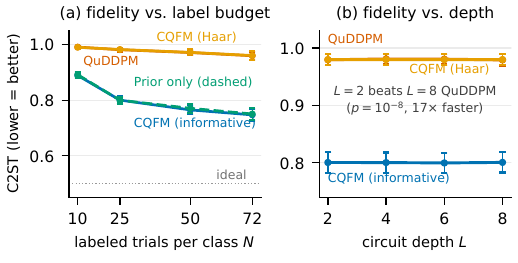}
\caption{(a) C2ST vs.\ label budget (mean $\pm$ 95\% CI over 9
subjects $\times$ 5 seeds). Informative-prior transport approaches
the data distribution as labels increase, while noise-initialized
transport stays almost perfectly distinguishable; the prior alone
(dashed) overlaps CQFM. (b) Fidelity does not change with circuit
depth ($N{=}25$, 9 subjects $\times$ 3 seeds).}
\label{fig:results}
\end{figure}

\begin{figure*}[t]
\centering
\includegraphics[width=\textwidth]{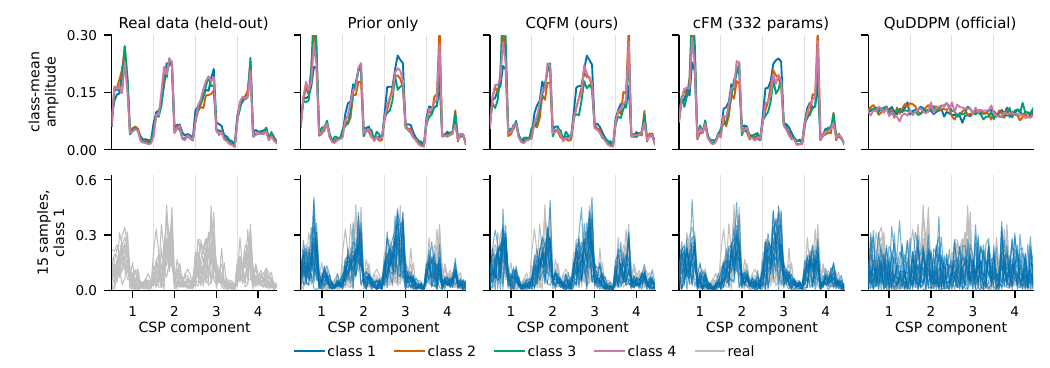}
\caption{What the models generate ($N{=}25$, subject S9---the median
subject by CQFM accuracy---seed 0). Top: class-mean spectra, where
four separated curves mean class structure is reproduced. Bottom: 15
generated samples of class 1 (colour) over 15 real ones (grey). Every
informative-prior model reproduces the CSP envelope and its class
structure; noise-initialized QuDDPM produces flat spectra.}
\label{fig:samples}
\end{figure*}

\textbf{Augmentation helps, and the prior decides by how much
(Table~\ref{tab:main}, Fig.~\ref{fig:results}a).} CQFM improves over
no augmentation by $+2.5$ points on average, on 9/9 subjects
($p{=}.004$, the floor for
$n{=}9$), and the improvement holds at every $N$. The gap to noise initialization is large: $+5.1$ points over the
official QuDDPM and $+5.4$ over the Haar-prior ablation (each 9/9,
$p{=}.004$). The ablation isolates the prior as the cause, since
circuit, budget, and training are identical. Fidelity metrics show
the same picture (Table~\ref{tab:main}, right): noise-initialized transport
remains almost perfectly distinguishable from real data (C2ST
$\approx$ 0.98), while informative-prior CQFM reaches 0.80 (9/9
subjects, $p{=}.004$); Fig.~\ref{fig:samples} shows the same contrast
directly in sample space. Conditional generators that start from the
informative prior, whether quantum or classical, all perform
similarly at the operating point.

\textbf{Depth does not compensate for the prior (Fig.~\ref{fig:results}b).} Sweeping $L \in
\{2,4,6,8\}$ (9 subjects, $N{=}25$, 3 seeds) leaves every method's
C2ST and LSD flat, and a \emph{2-layer} CQFM still dominates an
\emph{8-layer} QuDDPM (C2ST $-0.179$, 9/9 subjects, $p{=}.004$) at
$17\times$ lower training cost
(57\,s vs.~981\,s). Generation quality depends on where the transport starts
rather than on circuit capacity.

\textbf{Finite shots.} Reading amplitudes from computational-basis
samples at $N{=}25$, 1024 shots are indistinguishable from exact
(.411 vs.~.412); 256 shots cost 0.9 points, 5\% readout
depolarization 0.7.

\begin{figure}[t]
\centering
\includegraphics[width=\columnwidth]{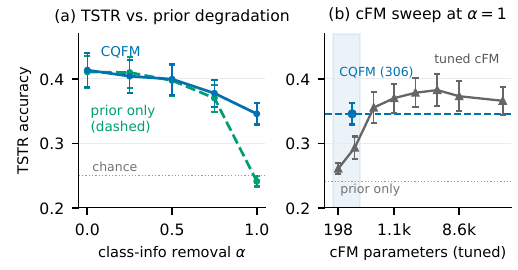}
\caption{(a) Prior-quality sensitivity ($N{=}25$): as class
information leaves the prior, the prior alone drops to chance TSTR
while CQFM recovers it through its conditioning. (b) Tuned classical FM
at $\alpha{=}1$ (best of four settings per budget). CQFM's 306
parameters fall between the 198- and 332-parameter models (shaded) and
beat both; classical catches up at about twice the size.}
\label{fig:prior}
\end{figure}

\textbf{Does the transport add value beyond its prior?
(Fig.~\ref{fig:prior}).} At the operating point it does not: the
prior alone already matches CQFM (Table~\ref{tab:main}). We degrade it with a class-information
shrinkage parameter $\alpha$ ($\mu_c(\alpha) = (1{-}\alpha)\mu_c +
\alpha\mu$, likewise $\sigma_c$); at $\alpha{=}1$ the prior carries no
label information, so any class structure must come from the
generator. Every conditional model starts from that same prior, and
the classical field is swept over eight sizes (198--33k) with the best
of four settings each (Fig.~\ref{fig:prior}b). Prior-only generation drops to chance TSTR (.241); CQFM recovers
.346, beating tuned classical FM at \emph{both} budgets bracketing its
306 parameters, .261 at 198 ($+8.5$, 9/9, $p{=}.004$) and .293 at 332
($+5.3$, 8/9, $p{=}.008$).
Classical FM catches up at about twice the size (.355 at
600) and plateaus at .37--.38, above CQFM's .346 but significant at
only one of eight budgets ($p{=}.04$), none after correction.

\textbf{Transferred priors: the misspecified case in practice.} The $\alpha$ knob is
synthetic; in practice a misspecified prior comes from \emph{other}
subjects. Pushing their trials through the target's encoder and
fitting the class Gaussian there displaces the prior by $1.3\times$ the
target's between-class distance on average.
Alone it drops TSTR from .411 to .299; CQFM recovers .370 ($+7.2$
points, 9/9, $p{=}.004$), landing within $1.1$ accuracy points of the
target's own prior ($p{=}.098$, n.s.). Against the size-matched
classical field CQFM gains $+3.5$ accuracy and $+4.4$ TSTR points (9/9
and 8/9, $p{\le}.008$) but \emph{loses} on distribution match (C2ST
$+0.090$, LSD $+0.57$, both $p{\le}.008$). Class geometry explains both:
scaled by within-class spread, class separation is 2.95 in real data,
1.37 in the transferred prior, 1.62 after classical FM and 2.22 after
CQFM (9/9 subjects, $p{=}.004$). The classical field polishes marginal
appearance and leaves the classes nearly as collapsed as it found
them.

\textbf{What the quantum channel contributes: nothing measurable.}
CQFM's advantage over an MLP velocity field could come from the
measured-ancilla channel or from its orthogonal near-identity
parameterization. We delete the ancilla and its measurement, changing
nothing else: a class-conditional orthogonal flow, 252 parameters at
$L{=}6$ and 336 at $L{=}8$, bracketing CQFM's 306. It matches CQFM on
all four metrics in both regimes and at both depths: TSTR .338 and
.343 against .346 at $\alpha{=}1$, and .369 against .370 with a
transferred prior, every comparison within noise (4--6 of 9 subjects,
$p{\ge}.13$). The measurement layer therefore buys nothing here. What
does pay is the parameterization: the same ablated flow, with fewer
parameters than CQFM, still beats the tuned MLP field by $+7.7$ TSTR
points at 198 (9/9, $p{=}.004$) and $+4.5$ at 332 (7/9, $p{=}.020$).
The gain is an inductive bias that happens to be expressible as a
quantum circuit, not an effect of measurement.

\section{Limitations and outlook}
At $6{+}1$ qubits the model can be simulated classically: each step
is a stochastic mixture of two $64\times64$ orthogonal maps. We therefore claim
no quantum computational advantage; the contribution is the model
design and the parameter-efficiency evidence above. Results are simulation-based: training uses exact branch
probabilities, and the noise analysis covers shots and readout
depolarization but not gate-level noise, while barren plateaus at
larger scale \cite{mcclean2018barren} remain open. Accuracies live in
the 64-dim feature space, well below raw-signal pipelines, and we
evaluate one EEG benchmark. Extending to ECG
\cite{qcganecg2023,hqdcgan2024,kritopoulos2026quantum} and to
raw-signal generation are natural next steps.

\section{Conclusion}
CQFM adds class conditioning to quantum flow matching and uses its
main freedom, the free choice of source ensemble: one shared circuit,
conditioned on flow time and class, transports a class-conditional
prior instead of noise. On BCI IV-2a it beats noise-initialized
quantum diffusion by a uniformly significant margin at a fraction of
the depth, and survives a prior transferred from other subjects. What
carries that is the conditional orthogonal parameterization: ablating
the measured ancilla costs nothing measurable.

\vfill\pagebreak

\section{Compliance with ethical standards}
This study uses only the publicly available, fully anonymized BCI
Competition IV-2a dataset \cite{tangermann2012review}; no new data
involving human or animal subjects were collected by the authors.

\bibliographystyle{IEEEbib}
\bibliography{refs}

\end{document}